\documentclass[10pt,twocolumn,letterpaper]{article}

\usepackage[utf8]{inputenc}
\usepackage[T1]{fontenc}
\usepackage{lmodern}
\usepackage{amsmath,amssymb,amsthm}
\usepackage{graphicx}
\usepackage{booktabs}
\usepackage{multirow}
\usepackage{array}
\usepackage{xcolor}
\usepackage{hyperref}
\usepackage{url}
\usepackage{enumitem}
\usepackage{algorithm}
\usepackage{algpseudocode}
\usepackage[font=small,labelfont=bf]{caption}
\usepackage[margin=0.75in]{geometry}
\usepackage{titlesec}
\usepackage{tikz}
\usetikzlibrary{shapes.geometric,arrows.meta,positioning,fit,backgrounds}

\hypersetup{colorlinks=true,linkcolor=blue!55!black,citecolor=green!45!black,
            urlcolor=blue!60!black}

\titleformat{\section}{\normalfont\bfseries\large}{\thesection.}{0.5em}{}
\titleformat{\subsection}{\normalfont\bfseries\normalsize}{\thesubsection}{0.5em}{}

\newcommand{\sysname}{GNoF}

\title{\bfseries GNoF: A Low-SM-Overhead GPU-Native NVMe-over-Fabrics\\
Completion Path for Data-Intensive GPU Workloads}

\author{%
  Lei Tian\thanks{Corresponding author: \texttt{tianlei01@nudt.edu.cn}}
  \quad Wenhao Gu\\
  National University of Defense Technology\\
  \texttt{\{tianlei01, guwenhao16\}@nudt.edu.cn}
}
\date{}

\begin{document}
\maketitle

\begin{abstract}
GPUs have become the primary compute engine for data-intensive workloads whose
datasets far exceed device memory and therefore reside on remote all-flash
storage. The conventional path lets the host CPU orchestrate every
NVMe-over-Fabrics (NVMe-oF) transaction and stage data through host memory,
incurring high CPU cost, repeated CPU--GPU synchronization, and redundant
data movement. GPU-initiated storage removes the CPU from the I/O path, but
existing designs leave two problems open on a fabrics target. Synchronous,
lock-step completion pins streaming multiprocessors (SMs) in spin-polling---
wasteful over a 10--100\,$\mu$s RDMA round trip---and current GPU-native
remote-storage systems concentrate their novelty in target-side firmware and
rely on per-command doorbells and periodic CQ polling on the client.

This paper attacks the \emph{client completion path} itself. We present
\sysname{}, a GPU-native NVMe-oF/RDMA data path in which GPU threads build
command capsules, post send work-queue elements, ring doorbells, and reap
completions entirely on the device while the CPU is retained only for
one-time connection setup, the admin queue, and a keep-alive thread. The
data path rests on three client mechanisms tuned for long fabric round
trips: a \emph{doorbell mover} that coalesces a burst's per-thread doorbells
into a single MMIO ring over its contiguous written prefix; a \emph{batched
CQ poller} built on a lock-free fast peek, one poll lock that drains the
whole ready batch, and monotonic per-slot completion counters that never
clear a flag and so cannot lose a wakeup; and a set-associative,
reference-pinned GPU \emph{page cache} with split fetch/wait pipelining and
orphan-fill takeover for on-demand remote paging. The design needs no target or SSD
firmware change and runs against an unmodified SPDK target on physical SSDs.

On a 200\,Gb/s InfiniBand fabric, \sysname{} sustains up to 1.77\,M 4\,KB
IOPS per client at \textbf{0.37\% host CPU} and \textbf{$\sim$1\% SM
throughput occupancy}, against 40\% host CPU for the kernel nvme-rdma
baseline and $\sim$100\% SM occupancy for lock-step polling. To put these
costs in context, we evaluate \sysname{} under identical workloads against
three existing baselines---kernel nvme-rdma, BaM over a BlueField-3 DPU
offload, and BaM on a local SSD---across SM occupancy, host/DPU resource
cost, interference immunity, three levels of I/O-induced GPU stall, and
graph/columnar applications. We report both the wins (selective column I/O
at 10--22$\times$, CPU-burn immunity) and the honest regime in which bulk
staging remains faster on a high-bandwidth fabric.
\end{abstract}

\section{Introduction}
\label{sec:intro}

GPU workloads such as graph analytics, recommender systems, and columnar data
processing share two traits: datasets of tens of GB to TB that cannot reside
in VRAM, and fine-grained, \emph{data-dependent} accesses whose targets are
only known after prior data is read. Serving these from remote flash is
standard in datacenters: a GPU node reaches an all-flash array (AFA) over
NVMe-oF/RDMA for capacity, sharing, and durability.

The established software path is CPU-centric. A kernel nvme-rdma initiator
(or an SPDK poller) owns the NVMe queues; the GPU must request data through
the CPU, which posts I/O, waits, and then copies the result into VRAM with
\texttt{cudaMemcpy}. Three costs follow.
\textbf{(1) Host CPU spend}: saturating the fabric with 4\,KB requests keeps
cores busy in the I/O stack.
\textbf{(2) CPU--GPU interaction overhead}: each transfer crosses a kernel
boundary and a synchronization point, and data detours through host memory
even when GPUDirect RDMA~\cite{gpudirectrdma} is enabled for the DMA itself;
GPUDirect Storage~\cite{gds} keeps direct DMA but likewise leaves
orchestration on the CPU.
\textbf{(3) I/O amplification}: analytic frameworks load coarse granules
(columns, tiles) to amortize per-request overhead, reading far more than the
application consumes.

GPU-initiated storage answers (BaM, libnvm, and filesystems built on them)
map NVMe doorbells into the GPU and let threads drive local PCIe SSDs
directly. They demonstrate that on-demand, cache-line-grained access can
eliminate amplification. But they target \emph{local} NVMe and adopt a
\textbf{synchronous lock-step model}: a thread submits a request and spins on
the completion queue until it returns. On a fabrics path the round-trip is an
order of magnitude longer than a local PCIe read, so na\"ive spin-polling
would occupy SMs for tens of microseconds per I/O with no useful work.
CoPilotIO~\cite{copilotio} observes exactly this waste and proposes moving
CQ polling back to
the CPU with hardware-barrier wakeups---reintroducing a host-side co-pilot
and a dependency on platform barrier/GDRCopy support; AGILE~\cite{agile}
instead dedicates SMs to I/O polling.

\paragraph{Question.} Can a GPU drive a \emph{remote} NVMe-oF target
directly, at high throughput, while keeping both the host CPU and the GPU SM
occupancy close to zero on the I/O path---without delegating polling back to
the CPU?

\sysname{} answers yes by reconstructing the verbs/libmlx5 data path in
device code over RDMA and replacing lock-step polling with event-friendly,
batched, lock-free queue management. The CPU is deliberately kept for the
slow/control plane (RDMA connection management, fabrics CONNECT, admin queue,
keep-alive), where its involvement is infrequent and off the critical path;
every steady-state I/O command is produced and consumed on the GPU.

\paragraph{Contributions.}
\begin{enumerate}[leftmargin=1.4em,itemsep=1pt,topsep=2pt]
\item A complete GPU-resident NVMe-oF/RDMA engine: device-side construction
of 64\,B NVMe command capsules with keyed SGLs targeting GPU memory, mlx5
SEND WQE posting, BlueFlame doorbells, and CQ parsing, byte-compatible with
the CPU libmlx5/provider contract so an \emph{unmodified, production} SPDK
target accepts the traffic with no target or SSD-firmware change
(\S\ref{sec:overview}--\S\ref{sec:datapath}).
\item Two queue mechanisms that let thousands of GPU threads share 64
queues without serializing on the NIC: a \emph{doorbell mover} that
coalesces a burst's per-thread doorbells into one MMIO ring, and a
\emph{batched CQ poller} with a lock-free fast peek and per-slot monotonic
completion counters, eliminating lost-wakeup races without flag clearing
(\S\ref{sec:submit}, \S\ref{sec:poll}).
\item A GPU page cache offering on-demand remote paging with non-blocking
fetch pipelines, reference-pinned eviction, and orphaned-fill \emph{takeover}
for progress under thread oversubscription---an on-demand caching layer
above the pre-registered buffer pool (\S\ref{sec:pcache}).
\item A thorough evaluation of \sysname{} that, to place its costs and
benefits in context, benchmarks it under identical workloads against three
existing reference architectures used as baselines---kernel nvme-rdma, BaM
over a BlueField-3 SNAP NVMe-oF offload, and BaM on a local SSD---measuring
SM occupancy, host/DPU resource cost, interference immunity, three levels of
I/O-induced GPU stall, and graph/columnar applications (\S\ref{sec:eval}).
The reference systems are prior work; our contribution here is the
controlled comparison and the resulting characterization, including the
regimes where GNoF wins and where bulk staging is preferable.
\end{enumerate}

\section{Background and Motivation}
\label{sec:bg}

\subsection{NVMe-oF/RDMA mechanics}
An RDMA-connected NVMe-oF initiator opens one RC queue pair (QP) per NVMe
I/O queue. A read or write is a SEND of a 64-byte NVMe command capsule whose
data pointer is a \emph{keyed SGL} naming a buffer and the MR remote key
(rkey); the target performs a one-sided RDMA WRITE (read) or READ (write)
against the initiator's registered memory, then SENDs a 16-byte completion
capsule. A submission queue (SQ) is a ring of WQEs in memory shared with the
HCA; progress requires writing the WQE, updating the doorbell record (DBrec),
and ringing the doorbell (a BlueFlame MMIO register on mlx5). Completions
land in a CQ ring whose entries carry an owner bit for lock-free consumer
synchronization; the consumer updates the CQ consumer-index DBrec. The command
ID (CID) names the slot the target's response refers to.

\subsection{Why the CPU stack is expensive}
We measure (\S\ref{sec:eval}) that the kernel nvme-rdma path at peak 4\,KB
throughput consumes $\approx$40\% of a host core and, because data lands in
host memory, requires an explicit transfer before the GPU can compute.
GPUDirect RDMA removes the host-memory detour for the DMA but the CPU still
orchestrates every command.

\subsection{Why local-GPU-storage designs do not transfer}
BaM maps the local NVMe controller's BAR0/doorbell to the GPU and exposes
\texttt{read\_data()}, which enqueues a command and \textbf{spins on the CQ in
place}. Local PCIe latency is low; a fabrics RTT ($\sim$10--100\,$\mu$s) makes
in-place spinning far more wasteful. Moreover, the lock-step contract (the
queue assumes submit$\to$poll within a grid that advances in lock-step) is
what enables the simple poll loop; our M4 experiments show BaM's device-side
async primitives are fragile under heterogeneous grids. \sysname{} keeps the
GPU-initiated model but redesigns the wait for an event/batched style suited
to long RTTs and many threads.

A na\"ive GPU-native fabrics initiator would carry over the per-command
pattern from local storage: each thread posts a WQE, rings the NIC doorbell,
and periodically polls its completion. Over a fabrics round trip this leaves
two scalability problems unaddressed---thousands of unsynchronized doorbells
serialize on the NIC MMIO region, and fine-grained completion locking turns
reaping into a convoy---while saying nothing about the SM cost of the wait.
\sysname{} therefore redesigns the client-side completion path itself,
around doorbell coalescing, batched lock-serialized CQ drainage, and an
on-demand GPU page cache (detailed in \S\ref{sec:datapath}), rather than
reusing a local-storage poll loop unchanged.

\subsection{Hardware/coherence constraints that shaped the design}
Three facts about discrete GPUs drive concrete mechanisms:
\begin{itemize}[leftmargin=1.2em,itemsep=1pt,topsep=2pt]
\item GPU L1 caches are not mutually coherent across SMs and device code
cannot invalidate a line; metadata shared across threads must use
\texttt{volatile}/atomic accesses, and memory produced by the NIC (RDMA
writes, CQEs) must be read with L1-bypass (\texttt{\_\_ldcg}) loads.
\item MMIO doorbell writes require a special store
(\texttt{st.mmio.relaxed.sys}); ordinary stores may not reach the device.
\item System-scope fences (\texttt{\_\_threadfence\_system()}) are required
before publishing WQEs/completions observed by another agent (HCA, peer
thread).
\end{itemize}

\section{Design Overview}
\label{sec:overview}

\begin{figure*}[t]
\centering
\resizebox{\textwidth}{!}{%
\begin{tikzpicture}[
  font=\small,
  box/.style={draw,rounded corners,align=center,minimum height=9mm,
              inner sep=4pt,fill=blue!4},
  gpubox/.style={box,fill=red!6},
  cpubox/.style={box,fill=green!6},
  net/.style={-{Latex[length=2mm]},thick},
  dashnet/.style={net,dashed},
]
\node[gpubox,minimum width=27mm] (app) {Application\\kernel};
\node[gpubox,minimum width=27mm,below=3mm of app] (pc) {GPU page cache\\(on-demand paging)};
\node[gpubox,minimum width=27mm,below=3mm of pc] (eng) {Device I/O engine\\capsule/WQE/doorbell/CQ};
\node[cpubox,minimum width=27mm,below=4mm of eng] (cpu)
  {Host control library\\CM connect, MR, keep-alive};
\node[box,minimum width=27mm,below=3mm of cpu,fill=orange!10] (nic)
  {RDMA NIC + peer memory};
\begin{scope}[on background layer]
  \node[draw,dashed,rounded corners,inner sep=7pt,fit=(app)(pc)(eng),
        label={[font=\small\itshape]north west:GPU (data plane)}] {};
  \node[draw,dashed,rounded corners,inner sep=7pt,fit=(cpu),
        label={[font=\small\itshape]north west:CPU (control plane)}] {};
\end{scope}
\node[box,right=22mm of nic,minimum width=24mm,fill=orange!10] (fab)
  {200\,Gb/s\\RDMA fabric};
\node[box,right=18mm of fab,minimum width=24mm] (tgt)
  {Unmodified\\SPDK NVMe-oF\\target};
\node[box,right=12mm of tgt,minimum width=20mm] (ssd)
  {NVMe\\SSD};
\draw[net] (nic)--(fab);
\draw[net] (fab)--(tgt);
\draw[net] (tgt)--(ssd);
\draw[dashnet,red!70!black] (eng.east) to[bend left=18]
  node[above,font=\scriptsize,text=black]{one-sided RDMA data into VRAM}
  (fab.north west);
\end{tikzpicture}}
\caption{GNoF splits the NVMe-oF stack by critical-path relevance. The GPU
owns every steady-state I/O operation (capsule build, WQE post, doorbell, CQ
poll, on-demand cache) while the CPU handles only one-time connection setup
and the admin/keep-alive path. The target is an unmodified SPDK exporter;
data is placed directly in VRAM by one-sided RDMA.}
\label{fig:arch}
\end{figure*}
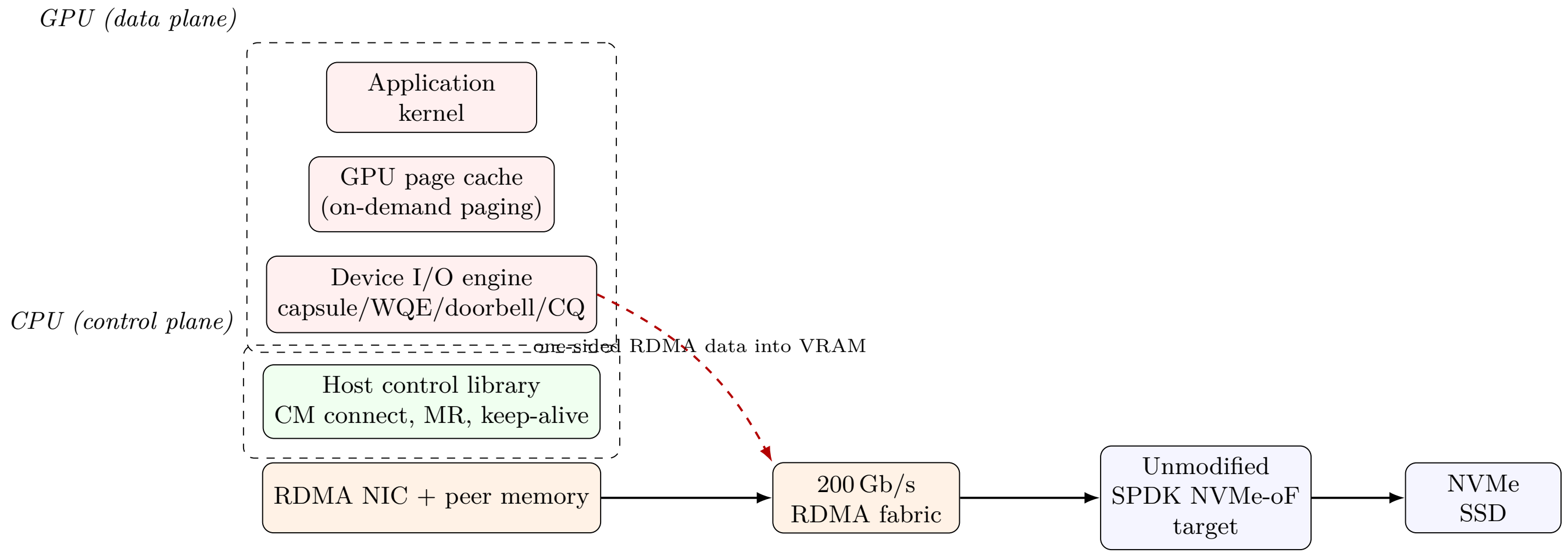

\sysname{} partitions the NVMe-oF stack by \emph{critical-path relevance},
not by historical CPU/GPU boundaries.

\paragraph{CPU (control plane, off the I/O critical path).}
At session open the host library (i) opens the HCA and protection
domain; (ii) RDMA-CM connects an admin QP and performs fabrics CONNECT,
Identify, and Set-Features (number of queues); (iii) creates one RC QP per I/O
channel with the correct private data, probing real mlx5 ring geometry via
\texttt{mlx5dv\_init\_obj}; (iv) allocates and registers a single GPU memory
pool (command/response buffers + data) with \texttt{ibv\_reg\_mr} over
\texttt{nvidia-peermem}; (v) posts the receive queue (RQ) of Recvs targeting
GPU response slots; (vi) \texttt{cudaHostRegister}s the mlx5 SQ/CQ/DBrec
buffers and, as I/O memory, the BlueFlame UAR page so the GPU can MMIO them;
and (vii) starts one keep-alive thread for the admin queue. After setup the
host blocks (not spins) and consumes effectively no steady-state CPU.

\paragraph{GPU (data plane).}
A \emph{channel} is one RC QP = one NVMe-oF I/O queue pair, with all
descriptors and a device state block resident in registered memory. Up to 64
channels spread concurrency; threads pick \texttt{chans[tid\,\%\,n]}. The
device API offers synchronous \texttt{gnof\_read/gnof\_write}, asynchronous
\texttt{gnof\_submit\_read/write} + \texttt{gnof\_wait}, and a page-cache API
(\texttt{gnof\_pc\_fetch/acquire/release}).

\paragraph{Target.}
An unmodified SPDK \texttt{nvmf\_tgt} exports a physical NVMe namespace over
RDMA. Because every \sysname{} descriptor mirrors what libmlx5 would produce,
no target-side change is required---a key deployability distinction from
designs that push functionality into SSD firmware or DPU agents.

\section{The GNoF Device Data Path}
\label{sec:datapath}

\subsection{Channel state and CID-as-slot identity}
Each channel holds the SQ/RQ/CQ rings, DBrecs, the BlueFlame register,
capsule/response arrays, geometry (\texttt{sq\_slots}, the HCA-padded
\texttt{sq\_ring\_cnt}, \texttt{rq\_stride}, \texttt{cq\_cnt}), MR keys, and a
device \texttt{state[]} block. Words 0--9 are monotonic counters
(\texttt{sq\_head}, doorbelled \texttt{sq\_db\_v}, send-CQE consumed, RQ
posted/done, CQ consumer index, BF toggle, and two locks); words 10 onward hold
per-slot \texttt{wrote[]}, monotonic \texttt{comp[]}, and status arrays.

The SQ slot index doubles as the NVMe CID, and a handle packs
\texttt{(target+1)<<32\,$|$\,slot}. Reusing the slot as CID lets the poller
map an incoming response's CID directly to its slot, and the biased encoding
guarantees a handle is never zero. A slot is claimable only while
\texttt{head$-$resp\_done < sq\_slots}, so a slot (and its capsule) is never
reused while its response is in flight, removing the send-CQE-lag race that
na\"ive ring wraparound causes.

\subsection{Submission and the doorbell mover}
\label{sec:submit}
the submit routine claims a slot with a single compare-and-swap on
\texttt{sq\_head}, fills the 64\,B capsule via \texttt{gnof\_build\_rw}
(keyed SGL with the pool rkey, zero-based NLB), writes the 32\,B mlx5 SEND
WQE at the \emph{true HCA ring position} (\texttt{cnt \& (sq\_ring\_cnt-1)},
because mlx5 pads the ring beyond the requested depth), then calls
\texttt{gnof\_sq\_commit}.

If every thread rang its own doorbell, thousands of MMIO writes and DBrec
updates would serialize on the BF register. The \textbf{doorbell mover}
instead has each thread set \texttt{wrote[cnt]=1} after a system fence, then
acquire a single \texttt{db\_lock}. The lock holder scans the contiguous
prefix of set \texttt{wrote[]} flags, clears them, advances
\texttt{sq\_db\_v}, and rings BlueFlame \textbf{once} for the whole batch
(DBrec update then a 64-bit BE BF MMIO of the control quadword with a
half-page toggle). This converts $N$ doorbells into roughly one per burst
while preserving WQE order. Algorithm~\ref{alg:submit} summarizes the path.

\begin{algorithm}[t]
\caption{GNoF submit + doorbell mover (per thread)}
\label{alg:submit}
\begin{algorithmic}[1]
\State // claim slot (== CID), bounded by in-flight guard
\Repeat $\text{head}-\text{resp\_done} < \text{sq\_slots}$
  \State $\text{cnt} \gets \text{CAS}(\text{sq\_head},\text{head},\text{head}+1)$
\Until{claimed}
\State $\text{slot} \gets \text{cnt}\,\&\,(\text{sq\_slots}-1)$; $\text{target}\gets\text{comp[slot]}$
\State build 64\,B keyed-SGL capsule; write 32\,B SEND WQE at ring position
\State \texttt{threadfence\_system()}; $\text{wrote[cnt]}\gets 1$
\State lock(\texttt{db\_lock}) \Comment{doorbell mover}
  \State advance over contiguous set \texttt{wrote[]} prefix; clear flags
  \State update DBrec; ring BlueFlame \textbf{once} for the whole prefix
\State unlock(\texttt{db\_lock})
\State \Return handle $(\text{target}+1){\ll}32 \;|\; \text{slot}$
\end{algorithmic}
\end{algorithm}

\subsection{Completion polling: lock-free peek, batched drain}
\label{sec:poll}
Waiters must reap both send CQEs (bookkeeping) and recv CQEs carrying the
16-byte NVMe response. A single poll lock protects CQ consumer state, but two
refinements keep it from becoming the bottleneck under dozens of waiting
threads:
\textbf{(1) Lock-free fast peek.} \texttt{gnof\_poll\_once} first checks the
head CQE's owner bit \emph{without} the lock; if nothing is valid it returns
immediately, so idle waiters never touch the lock word and cannot convoy
behind a thread with real work. A torn read is re-checked under the lock.
\textbf{(2) Batch drain under one acquisition.} Once the lock is held, the
poller processes \emph{every} currently-valid CQE in a loop, amortizing the
lock and system fences across all delivered responses rather than handing off
once per CQE.

For each recv CQE the poller reads the response (inline-data segment or the
RQ buffer indexed by the software recv tail, since non-SRQ RQ CQEs do not
carry the WQE index), extracts the CID, writes NVMe status into
\texttt{cplstat[slot]}, then \texttt{atomicAdd(comp[slot],1)}, re-posts the
consumed Recv, and finally updates the CQ consumer-index DBrec \emph{before}
publishing the new consumer counter (ordering that prevents a peer from
ringing a backwards CI at the wrap boundary).

Waiters spin on $\text{comp[slot]}>\text{target}$. Because \texttt{comp[]} is
monotonic and a waiter captures its target at claim time, \textbf{no
completion flag ever needs clearing}, which structurally eliminates
lost-wakeup races. The spin has a time-based ($\sim$10\,s) budget rather than
a fixed iteration count; expiry returns \texttt{0xffffffff} so the caller
surfaces an error instead of hanging. The same poll routine is invoked by
in-flight-full and page-cache waiters as \emph{helping}, so any thread makes
queue progress even while blocked on its own request.

\subsection{GPU page cache for on-demand remote paging}
\label{sec:pcache}
\texttt{gnof\_pc} carves fixed page slots from the registered pool and
implements a ways-way set-associative cache (Fibonacci-hashed set selection
from the LPN), with per-set spinlocks, global LRU stamp ticks, and per-slot
reference counts that \textbf{pin} referenced pages against eviction. Slot
state is EMPTY/FILLING/VALID.

\texttt{gnof\_pc\_fetch} is split from completion so a thread can pipeline
many misses: on a hit it pins and returns READY immediately; on a miss it
claims a victim (empty, else oldest unreferenced), marks FILLING, and submits
the fetch on a round-robin channel \emph{without waiting}, returning PENDING.
Other threads requesting the same LPN see FILLING and wait on the slot rather
than issuing duplicate I/O. \texttt{gnof\_pc\_wait} reaps the owner's handle,
or---if the owner appears wedged for a takeover grace ($\sim$100\,ms)---reaps
the recorded in-flight handle itself (safe because \texttt{gnof\_wait} is
valid from any thread and an already-completed handle returns immediately),
publishing VALID only while the slot still carries the same fill. With no
free way, fetch returns RETRY instead of blocking while holding pins,
preventing the congestion collapse of deep batch pipelines; callers drain
outstanding requests to release pins and retry. Every metadata word is
accessed volatile/atomically, and NIC-filled pages are consumed through
\texttt{\_\_ldcg} to bypass stale L1 lines. This cache supplies the
BaM-style on-demand abstraction \emph{over a remote target}, with the
asynchronous semantics a long RTT demands.

\subsection{Why SM occupancy stays low}
Unlike lock-step \texttt{read\_data}, a \sysname{} waiter does not occupy an
SM in a tight compute loop: submissions are amortized by the mover,
completions are batch-reaped, and waiting is on memory flags with helping.
Our measurements (\S\ref{sec:eval}) show SM throughput occupancy of
0.03--1.07\% and warps-active capped at 16.67\%, against $\sim$100\% SM issue
for BaM-style polling---the direct consequence of replacing spin-polling with
event/batched completion.

\section{Implementation}
\label{sec:impl}
\sysname{} is implemented in CUDA C++ for an Ada-generation GPU
(compute capability 8.9) plus a user-space host library over rdma-core/verbs
and mlx5 Direct Verbs. The device data path is a header-only library linked
into application kernels; the host library manages RDMA CM/DevX connections,
memory-region registration, host registration of HCA queue objects, and the
single keep-alive thread. NVMe-oF capsule and WQE layouts follow the NVMe-oF
RDMA transport binding and the verbs provider contract, and are
cross-checked against the kernel nvme-rdma initiator. A single memory
region covers command capsules, response buffers, and data; a peer-memory
kernel module~\cite{peermem} enables one-sided NIC DMA directly into VRAM.
The prototype is on the order of a few thousand lines of CUDA/C++. The block
microbenchmark, the BFS/CC graph engine, the column-scan analytic, and the
vector-add kernel, together with the BaM baseline, all reuse the same
registered-pool and 4\,KB page geometry. Table~\ref{tab:impl} summarizes the
main tunables; their values are the defaults used throughout the evaluation.

\begin{table}[t]
\centering\small
\caption{Principal prototype parameters and their default values.}
\label{tab:impl}
\resizebox{\columnwidth}{!}{%
\begin{tabular}{ll}
\toprule
Parameter & Default \\
\midrule
I/O channels (RC QPs = NVMe-oF queues) & 64 \\
SQ slots per channel (outstanding cap) & 256 \\
CQ entries per channel & 1024 \\
Page-cache page size & 4\,KB \\
Page-cache associativity & set-associative, 8 ways \\
Doorbell coalescing & per contiguous WQE prefix \\
\bottomrule
\end{tabular}}
\end{table}

\section{Evaluation}
\label{sec:eval}

\noindent\textbf{Key results at a glance.}
\begin{itemize}[leftmargin=1.3em,itemsep=1pt,topsep=3pt]
\item \textbf{Near-zero host and SM cost}: 0.37\% host CPU and $\sim$1\% SM
throughput occupancy at peak, vs.\ 40\% host CPU for the kernel path and
$\sim$100\% SM issue for lock-step polling.
\item \textbf{Full fabric throughput}: up to 1.77\,M 4\,KB IOPS per client
with no target modification.
\item \textbf{Interference immunity}: normalized throughput stays $\approx$1.0
while 36 host cores are saturated; a DPU-only bottleneck is exposed at 8
burned DPU Arm cores.
\item \textbf{Async overlap}: 2.60$\times$ speedup from compute/I/O overlap
with zero failed runs, where lock-step device primitives cannot run at all.
\item \textbf{Selective analytic I/O}: 10--22$\times$ on cold column scans
with bit-identical results, alongside an honest boundary where bulk staging
wins on the large graph.
\end{itemize}

\subsection{Setup and schemes}
Our testbed consists of a GPU client node and a remote storage target
connected by a 200\,Gb/s InfiniBand fabric. The client hosts an Ada-generation
datacenter GPU (46\,GB VRAM, compute capability 8.9) and a BlueField-3
data-processing unit; the target attaches high-endurance enterprise NVMe
SSDs exported by an unmodified SPDK NVMe-oF target. The GPU and NIC sit across
a NUMA boundary (the \emph{SYS} topology class in NVIDIA's topology matrix);
we report this rather than assume near-switch peer-to-peer placement. A
peer-memory module enables NIC-to-VRAM DMA. Where applicable the schemes
share one target namespace and are aligned by \textbf{total in-flight
requests $C$} (not thread or core count); all measurements use direct I/O,
strictly one scheme at a time, and three repetitions. We compare four
schemes:
\begin{itemize}[leftmargin=1.2em,itemsep=1pt,topsep=2pt]
\item \textbf{P1 --- kernel nvme-rdma + fio/io\_uring}, data to host memory
(CPU-centric baseline);
\item \textbf{P2 --- \sysname{}}, GPU-driven RDMA into VRAM (this paper);
\item \textbf{P3 --- BaM over BF3 SNAP}: GPU drives a SNAP-emulated NVMe
controller whose DPU Arm initiator connects to the target (queue depth capped
by SNAP MQES=64);
\item \textbf{P4 --- BaM on a local SSD}: GPU P2P to a local NVMe controller
(no network).
\end{itemize}

\subsection{SM occupancy (the polling question)}
\begin{figure}[t]
\centering
\includegraphics[width=\columnwidth]{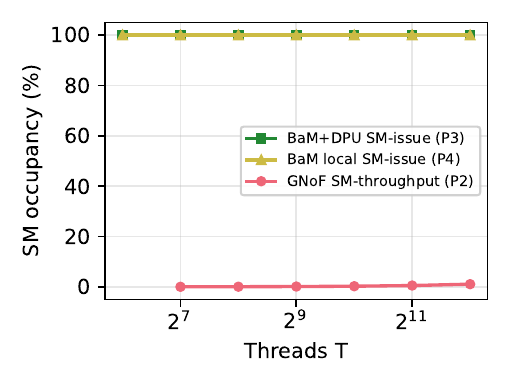}
\caption{SM occupancy vs.\ thread count. BaM paths sit at $\sim$100\% SM
issue (spin-polling) at every concurrency, while GNoF's event/batched
completion keeps SM throughput near zero.}
\label{fig:sm}
\end{figure}

With Nsight (ncu for \sysname{}; nsys GPU metrics for BaM, since ncu cannot
replay its BAR-side-effect kernels) across thread counts 64--4096: BaM
(P3/P4) holds SM-issue/GR-active $\approx$100\% at every count---pure
spin-polling. \sysname{} measures sm\_throughput 0.03$\to$1.07\% and
warps-active $\leq$16.67\%. P1 has no GPU I/O kernel (0\%). This is the
central resource-efficiency result: GPU-native I/O need not cost the SM
(Table~\ref{tab:e1}, Figure~\ref{fig:sm}).

\begin{table}[t]
\centering\small
\caption{SM occupancy vs.\ thread count $T$ (representative rows).}
\label{tab:e1}
\resizebox{\columnwidth}{!}{%
\begin{tabular}{lcccc}
\toprule
$T$ & P2 SM-throughput\% & P2 warps-active\% & P3/P4 SM-issue\% & P1 \\
\midrule
128  & 0.03 & 8.33  & $\approx$100 & 0 \\
512  & 0.13 & 16.67 & $\approx$100 & 0 \\
2048 & 0.53 & 16.67 & $\approx$100 & 0 \\
4096 & 1.07 & 16.67 & $\approx$100 & 0 \\
\bottomrule
\end{tabular}}
\end{table}

\subsection{Throughput and resource cost}
\begin{figure}[t]
\centering
\includegraphics[width=\columnwidth]{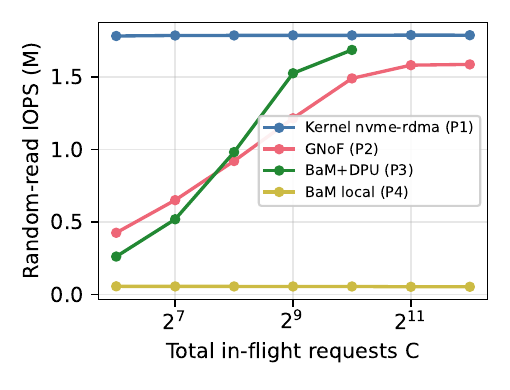}
\caption{4\,KB random-read IOPS vs.\ total in-flight requests $C$. The three
fabric-attached schemes reach 1.2--1.8\,M IOPS; the local BaM read path
plateaus low regardless of concurrency.}
\label{fig:throughput}
\end{figure}

\textbf{4\,KB throughput.} P1/P2/P3 all reach 1.2--1.8\,M IOPS as $C$ scales
64$\to$4096; P1 saturates as early as $C{=}64$ while \sysname{} and P3 ramp
with concurrency. P4 reads plateau at only $\sim$55\,K (random) /
$\sim$126\,K (sequential) IOPS independent of $C$---a same-drive fio test
(1.76\,M IOPS) verifies this is BaM's lock-step read path, not the device
(7\% of capability); its writes scale normally to 1.17\,M. P3 cannot exceed
$C{=}1024$ because the SNAP controller's MQES=64 rejects larger data-plane
fan-out (FAIL, not a missing sample). \sysname{} peaks at $\sim$1.77\,M reads
/ $\sim$1.0\,M writes at essentially zero reported CPU
(Figure~\ref{fig:throughput}).

\begin{figure}[t]
\centering
\includegraphics[width=0.92\columnwidth]{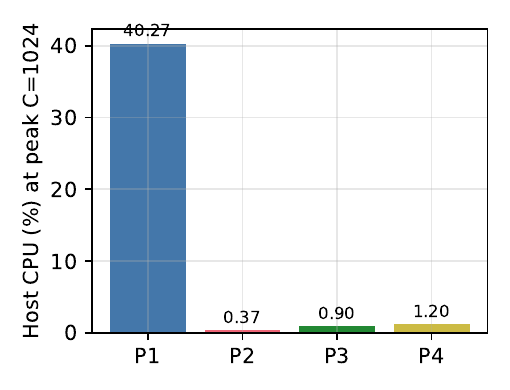}
\caption{Host CPU consumption at peak concurrency. GNoF removes the CPU from
the I/O path; the remaining cost of GPU/DPU-initiated schemes is on the
device rather than the host.}
\label{fig:cpu}
\end{figure}

\textbf{Resource cost at $C{=}1024$} (Table~\ref{tab:m2},
Figure~\ref{fig:cpu}). The cost does not
disappear under offload---it moves: P1 spends host cores, P4 spends GPU spins,
P3 additionally spends a quarter of the DPU Arm.

\begin{table}[t]
\centering\small
\caption{Resource cost at peak concurrency $C{=}1024$.}
\label{tab:m2}
\resizebox{\columnwidth}{!}{%
\begin{tabular}{lcccc}
\toprule
Resource & P1 & P2 \sysname{} & P3 BaM+DPU & P4 BaM local \\
\midrule
Host CPU \%            & 40.3  & \textbf{0.37} & 0.9  & 1.2 \\
GPU util avg/peak \%   & 0/0   & 19/100        & 75/100 & 94/100 \\
Peak VRAM (MiB)        & 0     & 8702          & 552  & 552 \\
DPU Arm \%             & ---   & ---           & 24.8 & --- \\
\bottomrule
\end{tabular}}
\end{table}

\subsection{Single-I/O latency}
At strict in-flight=1 with $\geq$500 requests per point and full percentiles,
P1 and \sysname{} show comparable end-to-end RTT ($\sim$98\,$\mu$s 4\,KB read)
but \sysname{} uses 0.2\% host CPU vs 17.6\% and has a lower p99 (135 vs
197\,$\mu$s). BaM numbers from the stock benchmark are batch-amortized
(\texttt{elapsed/ops} over 64 threads), not strict $C{=}1$ RTT, and are
reported in separate columns with an explicit caveat; a same-caliber BaM
single-flight latency test is future work (the stock tool emits no percentile
distribution).

\subsection{Interference immunity}
\begin{figure}[t]
\centering
\includegraphics[width=\columnwidth]{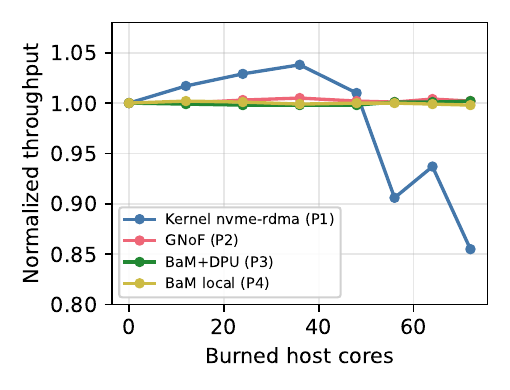}
\caption{Normalized throughput under host-core contention. GPU/DPU-initiated
paths stay near 1.0 while host cores are saturated; the CPU-centric path
degrades once contention is high enough.}
\label{fig:interf}
\end{figure}

Burning host cores with pinned \texttt{while(1)} loops (0$\to$72 cores):
\sysname{}/P3/P4 retain normalized throughput $\approx$1.00 through 36 burned
cores because their I/O bypasses the host CPU; P1 degrades to 0.84--0.94 only
past 56 cores. Burning the \textbf{DPU Arm} (P3 only) leaves throughput flat
through 4 cores but drops it to 0.59 at 8---exposing the SNAP initiator as
P3's structural bottleneck, a dimension invisible in all-host studies
(Figure~\ref{fig:interf}).

\subsection{I/O-induced GPU stalls (after CoPilotIO's taxonomy)}
\emph{Intra-warp overlap:} with an async submit/wait pipeline, \sysname{}
reaches 2.60$\times$ speedup over a synchronous baseline across compute/IO
ratios (36/36 zero-failure runs); P1 reaches 2.25$\times$ but pays a
$\sim$139\,$\mu$s per-round kernel-launch cost. BaM's device async primitives
race (Xid43) and cannot run this test---a structural limitation, reported as
such.
\emph{Inter-warp contention} (FP32 issue-bound): P1 stalls compute warps 0\%
even at 1.79\,M IOPS; \sysname{} 6.8$\to$12.3\%; BaM 7$\to$15.1\% (same regime
as CoPilotIO's reported $\sim$20\%).
\emph{Inter-SM contention:} P1 retains 1.0 bandwidth under up to 48 copy SMs
(host-memory/NIC path is physically separate from VRAM); \sysname{} stays
within 5--8\% loss through 32 SMs but drops to 0.227 retention at 48 when the
DRAM subsystem saturates and CQ polling is starved. BaM cannot form the
heterogeneous grid (lock-step assumption), documented with an exclusion
chain rather than silently omitted.

\subsection{Applications}
\textbf{Graph BFS/CC (R-MAT).} Reachable-vertex counts match across schemes.
On the small graph (2.13\,GB) \sysname{} fully-resident is fastest (1975\,ms
vs staged 2567 vs host 3998). On the large graph (8.5\,GB) bulk \emph{staging}
wins (9873\,ms) over \sysname{} (11870\,ms) and host (16798\,ms): on a
965\,MB/s fabric the bulk-load tax has fallen enough that coarse staging
beats many fine on-demand reads---an honest regime boundary. A cache-size
sweep shows BFS time falling monotonically 61.3\,s$\to$9.65\,s as the page
cache grows 1$\to$16\,GiB.
\textbf{NYC taxi column scan} (38.3\,M rows, 2.85\,GB image, 6 queries):
results are bit-identical to an independent NumPy reference; cold \sysname{}
runs take 224--804\,ms against a fixed $\sim$5\,s full-image bulk load for
T---\textbf{10--22$\times$} from selective column I/O.
\textbf{Vector add} (write-heavy, regular): \sysname{} peaks at 863\,MiB/s at
$T{=}512$ but does not beat the best staged pipeline (1.56\,s vs 3.56\,s);
here the value is zero CPU / no tiling / whole-array GPU naming, not raw
throughput. Table~\ref{tab:apps} summarizes the application-level picture.

\begin{table}[t]
\centering\footnotesize
\caption{Application-level results (lower time is better; speedup is GNoF vs.\
the bulk-loading T baseline).}
\label{tab:apps}
\resizebox{\columnwidth}{!}{%
\begin{tabular}{llll}
\toprule
Workload & Metric & GNoF & Staged / T \\
\midrule
BFS, 2.13\,GB & time (ms) & \textbf{1975} & 2567 / 3998 \\
BFS, 8.5\,GB & time (ms) & 11870 & \textbf{9873} / 16798 \\
Taxi scan & cold (ms) & \textbf{224--804} & $\sim$5000 \\
Taxi selective & speedup & \textbf{10--22$\times$} & 1$\times$ \\
Vector add & time (s) & 3.56 & \textbf{1.56} \\
\midrule
Correctness & \multicolumn{3}{l}{bit-identical to NumPy; BFS reach-set matches} \\
\bottomrule
\end{tabular}}
\end{table}

Application-level BaM numbers are the principal remaining gap (BaM ships
BFS/CC/vector benchmarks but they were not integrated into this campaign).

\subsection{Limitations and honest scope}
(1) \sysname{} is a client-side library: it does not implement multi-client
coordination, access control, or replication. (2) GPU$\leftrightarrow$NIC is
SYS topology; near-switch P2P numbers may differ. (3) M4a/M4c quantify BaM
library limits rather than a fundamental GPU property. (4) Same-caliber
single-flight BaM latency and application-level BaM/CoPilotIO comparisons are
future work. (5) On regular bulk writes \sysname{} is throughput-neutral vs
staging; its advantage is selectivity and resource cost, not peak bulk
bandwidth.

\section{Related Work}
\paragraph{GPU-initiated local storage.}
BaM~\cite{bam} and libnvm~\cite{libnvm} map NVMe doorbells to GPUs for local
SSDs with synchronous polling and on-demand caches.
\sysname{} adopts GPU initiation and on-demand
paging but rebuilds the verbs path for fabrics with batched/event completion
to avoid SM spinning.
\paragraph{CPU-assisted GPU I/O.}
CoPilotIO~\cite{copilotio} splits SQ/CQ across GPU/host memory and uses CPU
co-polling plus
\texttt{cuda::barrier} wakeups; AGILE~\cite{agile} reserves SMs for polling. \sysname{}
keeps polling on the GPU but removes its cost through batching and monotonic
completion, avoiding host co-pilots, GDRCopy, and hardware-barrier platform
dependencies.
\paragraph{GPUDirect storage/networking.}
GDS~\cite{gds} and GPUDirect RDMA~\cite{gpudirectrdma} enable direct DMA but
retain CPU orchestration;
\sysname{} moves orchestration onto the device.
\paragraph{GPU-native remote all-flash arrays.}
A concurrent GPU-native all-flash-array
system~\cite{gnstor} targets the \emph{storage-target} boundary: alongside a
GPU-oriented NVMe-oF initiator it contributes a decentralized engine that
moves access control and the logical-to-physical map into SSD firmware for
multi-client, fault-tolerant sharing, and is evaluated largely with an
emulated SSD. Its initiator and \sysname{} share the high-level
GPU-initiation pattern---channels, compare-and-swap synchronization,
pre-registered GPU memory pools, and a CPU reserved for connection setup and
keep-alive---but the two efforts place their emphasis differently. That work
centers on target-side coordination and a multi-level GPU memory allocator,
whereas \sysname{} is a single-client, target-agnostic data path whose
contribution is the \emph{client completion mechanics} for long fabric RTTs:
doorbell coalescing across a burst, batched CQ drainage with monotonic
per-slot completion counters, and an on-demand GPU page cache with
orphan-fill takeover, together with an explicit treatment of discrete-GPU
non-coherence. Correspondingly, \sysname{} needs no target or firmware change
and runs against an unmodified SPDK target~\cite{spdk} over the NVMe-oF RDMA
transport~\cite{nvmespec} with standard mlx5 verbs~\cite{mlx5verbs}, and our
evaluation quantifies the SM cost of the wait strategy
($\sim$1\% vs.\ $\sim$100\% spin-polling) through a four-scheme comparison
rather than a two-baseline AFA study. GPU filesystems
(GoFS~\cite{gofs}, GeminiFS~\cite{geminifs}) layer on GPU block engines and
are complementary; RAPIDS~\cite{rapids} motivates GPU-resident analytic
engines that consume such storage.

\section{Conclusion}
\sysname{} shows that a GPU can drive a remote NVMe-oF target end-to-end---
capsules, WQEs, doorbells, completions, and an on-demand page cache---while
keeping the host CPU at 0.37\% and SM throughput occupancy around one percent,
matching the throughput of CPU-centric paths without their resource cost and
avoiding the SM spin of synchronous GPU-storage designs. Its batched,
monotonic, coherence-aware queue mechanics are the enabling ingredients. The
accompanying four-scheme study delineates where GPU-native fabrics access is
decisive (selective analytic I/O, interference immunity) and where simple
bulk staging remains pragmatic.

\appendix
\section{Reproducibility Considerations}
\label{app:repro}
All reported numbers are aggregate statistics (means over repetitions, peak
throughput, occupancy, and normalized ratios) rather than raw traces. The
four schemes are evaluated on one testbed under one common workload
definition, aligned by total in-flight requests $C$, with direct I/O and
three repetitions per point; each scheme runs in isolation. The device-side
mechanisms are fully specified by Algorithm~\ref{alg:submit} and
\S\ref{sec:datapath} (state-block layout, owner-bit validity, monotonic
completion counters, and the page-cache state machine), and require only
standard CUDA atomics, system-scope fences, MMIO stores, and the documented
NVMe-oF RDMA verbs contract to re-implement. Experiments exercise only
dedicated data drives; no operating-system disk is ever re-bound. A
reference implementation and the per-point measurement dataset may be
released by the authors separately.

\section{Mechanism Summary}
Table~\ref{tab:mech} maps each design element of \S\ref{sec:datapath} to the
coherence or concurrency problem it resolves; this is the conceptual index
that an independent re-implementation would follow, without reference to any
proprietary source layout.
\begin{table}[h]
\centering\small
\caption{Mechanism-to-problem map (design-level, no source disclosure).}
\label{tab:mech}
\resizebox{\columnwidth}{!}{%
\begin{tabular}{p{0.46\columnwidth}p{0.54\columnwidth}}
\toprule
Mechanism & Problem addressed \\
\midrule
Doorbell mover with prefix coalescing & One MMIO per burst instead of one per thread; preserves WQE order \\
Lock-free peek + single poll lock + batched drain & Avoids poll-lock convoys; amortizes fences over a CQE batch \\
Monotonic per-slot completion counters & No flag clearing; structurally free of lost-wakeup races \\
Slot-as-CID + in-flight guard & Direct response-to-slot mapping; no capsule reuse while in flight \\
Time-bounded spin with queue helping & Liveness under contention; bounded failure instead of hangs \\
Set-associative, reference-pinned cache & Prevents eviction of in-use pages; LRU within a set \\
Fetch/wait split + orphan-fill takeover & Miss pipelining and progress if a filling thread deschedules \\
L1-bypass loads for NIC-written memory & Correctness under non-coherent GPU L1 \\
System-scope fences before publish & Orders WQE/cache state observed by HCA and peer threads \\
\bottomrule
\end{tabular}}
\end{table}

\bibliographystyle{plain}
\bibliography{refs}

@inproceedings{bam,
  title     = {{GPU-Initiated} On-Demand High-Throughput Storage Access in the {BaM} System Architecture},
  author    = {Qureshi, Zaid and Mailthody, Vikram Sharma and Gelado, Isaac and
               Min, Seungwon and Masood, Amna and Park, Jeongmin and Xiong, Jinjun and
               Newburn, C. J. and Vainbrand, Dmitri and Chung, I-Hsin and
               Garland, Michael and Dally, William and Hwu, Wen-mei},
  booktitle = {Proc. 28th Int'l Conf. on Architectural Support for Programming
               Languages and Operating Systems (ASPLOS), Volume 2},
  pages     = {325--338},
  year      = {2023},
  doi       = {10.1145/3575693.3575748},
  note      = {Vancouver, BC, Canada; ACM ISBN 978-1-4503-9916-6}
}

@inproceedings{copilotio,
  title     = {{CoPilotIO}: {CPU} as a Co-pilot for {GPU} {I/O} to Free {GPU} Compute},
  author    = {Chen, Guanyi and Chen, Qi and Yin, Shu and Zhang, Jian},
  booktitle = {Proc. 20th USENIX Symp. on Operating Systems Design and
               Implementation (OSDI)},
  year      = {2026},
  note      = {Seattle, WA, USA; ISBN 978-1-939133-55-7}
}

@misc{gnstor,
  title        = {{GNStor}: Design of {GPU-Native} High-Performance Remote All-Flash Array},
  author       = {Yi, Shushu and Wu, Wenbo and Chen, Guoci and Zhu, Junrong and
                  Liang, Shengwen and Bo, Mao and Huan, Chenying and Tian, Chen and Zhang, Jie},
  howpublished = {arXiv preprint arXiv:2606.04908},
  year         = {2026}
}

@misc{libnvm,
  title        = {{libnvm}: A Userspace Library for {GPU} Access to {NVMe} Devices},
  author       = {{libnvm contributors}},
  howpublished = {Open-source software project}
}

@inproceedings{agile,
  title     = {{AGILE}: Lightweight and Efficient Asynchronous {GPU-SSD} Integration},
  author    = {Yang, Zhuoping and Zhuang, Jinming and Chen, Xingzhen and
               Jones, Alex K. and Zhou, Peipei},
  booktitle = {Proc. Int'l Conf. for High Performance Computing, Networking,
               Storage and Analysis (SC)},
  pages     = {1028--1042},
  year      = {2025},
  note      = {St. Louis, MO, USA}
}

@manual{gds,
  title        = {{NVIDIA GPUDirect Storage}: A Direct Path Between Storage and {GPU} Memory},
  author       = {{NVIDIA Corporation}},
  howpublished = {Technology documentation},
  year         = {2023}
}

@manual{gpudirectrdma,
  title        = {{NVIDIA GPUDirect RDMA}},
  author       = {{NVIDIA Corporation}},
  howpublished = {Technology documentation},
  year         = {2023}
}

@manual{nvmespec,
  title        = {{NVM Express} Base Specification and {NVMe-oF} {RDMA} Transport},
  author       = {{NVM Express, Inc.}},
  howpublished = {Specification}
}

@manual{spdk,
  title        = {{SPDK}: Storage Performance Development Kit},
  author       = {{Storage Performance Development Kit community}},
  howpublished = {Open-source storage framework}
}

@manual{mlx5verbs,
  title        = {{RDMA Verbs and Direct Verbs (mlx5dv)} Programming Guide},
  author       = {{NVIDIA Networking}},
  howpublished = {Programming manual}
}

@inproceedings{gofs,
  title     = {Managing Scalable Direct Storage Accesses for {GPUs} with {GoFS}},
  author    = {Li, Shaobo and Zhou, Yirui Eric and Xue, Yuqi and Xu, Yuan and Huang, Jian},
  booktitle = {Proc. 31st ACM Symp. on Operating Systems Principles (SOSP)},
  pages     = {979--995},
  year      = {2025},
  note      = {Seoul, Republic of Korea}
}

@inproceedings{geminifs,
  title     = {{GeminiFS}: A Companion File System for {GPUs}},
  author    = {Qiu, Shi and Liu, Weinan and Hu, Yifan and Yan, Jianqin and
               Shen, Zhirong and Yao, Xin and Chen, Renhai and Zhang, Gong and Zhang, Yiming},
  booktitle = {Proc. 23rd USENIX Conf. on File and Storage Technologies (FAST)},
  pages     = {221--236},
  year      = {2025},
  note      = {Santa Clara, CA, USA}
}

@misc{rapids,
  title        = {{RAPIDS}: {GPU}-Accelerated Data Analytics Libraries},
  author       = {{NVIDIA Corporation}},
  howpublished = {Open-source software},
  year         = {2023}
}

@manual{peermem,
  title        = {{nvidia-peermem}: {GPUDirect RDMA} Peer-Memory Kernel Module},
  author       = {{NVIDIA Corporation}},
  howpublished = {Kernel driver documentation}
}
\end{document}